%% file: main.tex
\documentclass[final,finalrunningheads,a4paper]{llncs}

\input{styles.tex}
\input{macro.tex}

\newtoggle{anonymous}
\togglefalse{anonymous}

\makeatletter
\renewcommand\paragraph{\@startsection{paragraph}{4}{\z@}%
  {2.25ex \@plus 1ex \@minus .2ex}%
  {-0.75em}%
  {\normalfont\normalsize\bfseries}}
\makeatother

\begin{document}

\title{Bitcoin covenants unchained}

\iftoggle{anonymous}{}{
\author{Massimo Bartoletti\inst{1}, Stefano Lande\inst{1}, Roberto Zunino\inst{2}}
\authorrunning{Bartoletti et al.}

\institute{
Universit\`a degli Studi di Cagliari, Cagliari, Italy
\and 
Universit\`a degli Studi di Trento, Trento, Italy}
}

\maketitle

\input{abstract.tex}

\input{intro.tex}

\input{bitcoin.tex}

\input{covenants.tex}

\input{examples.tex}
  \input{crowdfunding.tex}
  \input{non-fungible-token.tex}
  \input{vault.tex}

  \input{ponzi.tex}
  \input{kotet.tex}

\input{implementation.tex}

\input{bitml-covenants.tex}

\input{conclusions.tex}

\input{ack.tex}

\bibliographystyle{splncs04}
\bibliography{main}

\end{document}

%% file: styles.tex
\usepackage[utf8]{inputenc}
\usepackage{color}
\usepackage[usenames,dvipsnames]{xcolor}

\usepackage{centernot}
\usepackage{hhline}

\usepackage{bm} % for bold symbols

\usepackage{siunitx}
\usepackage{textcomp}
\usepackage{caption}
\usepackage{pst-func}
\usepackage{amssymb}
\usepackage{graphicx}
\usepackage{paralist}
\usepackage{environ}
\usepackage{pgfplots}
\usepackage{tikz}
\usetikzlibrary{shadows}
\usepackage{mathtools}
\usepackage{array,multirow}
\usepackage{algorithm}
\usepackage[hyphens]{url}
\usepackage{cite}
\usepackage{pgfplotstable}
\usepackage{threeparttable}

\usepackage{etoolbox}
\usepackage{arydshln}

\usepackage{xspace} % Smart spaces after commands
 
\usepackage{dirtytalk} % for \say command
\usepackage{stmaryrd} % for llbracket and rrbracket\textbf{\textbf{}}

\usepackage[inline,shortlabels]{enumitem} 
\newlist{inlinelist}{enumerate*}{1}
\setlist*[inlinelist,1]{%
  label=(\roman*),
}

\usepackage{xifthen}  % Extended conditional commands
\usepackage{ifthen}

\usepackage{nicefrac}

\usepackage[hidelinks]{hyperref}
\usepackage{cleveref}
\usepackage{bigints}

\usetikzlibrary{pgfplots.dateplot}
\usetikzlibrary{matrix,chains,positioning,decorations.pathreplacing}
\usetikzlibrary{patterns,calc,fit,arrows}
\usetikzlibrary{shapes.geometric}

\usepackage[final,nomargin,inline,index]{fixme} % Simplified management of FIXME's
\fxusetheme{color}

\FXRegisterAuthor{bart}{anbart}{\color{magenta} {\underline{bart}}}
\FXRegisterAuthor{tizi}{antizi}{\color{red} {\underline{tizi}}}
\FXRegisterAuthor{ste}{anste}{\color{blue} {\underline{ste}}}
\FXRegisterAuthor{nico}{annico}{\color{ForestGreen} {\underline{nico}}}

\hypersetup{
  breaklinks   = true,
  colorlinks   = true, %Colours links instead of ugly boxes
  urlcolor     = blue, %Colour for external hyperlinks
  linkcolor    = blue, %Colour of internal links
  citecolor    = red   %Colour of citations
}
\hypersetup{final} % Needed to generate hyperlinks even in draft mode (ERROR IN MAC)

\usepackage{listings}
\definecolor{listingBG}{HTML}{FFFFCB}%
\definecolor{listingFrame}{HTML}{BBBB98}%
\definecolor{listingLineno}{rgb}{0.5,0.5,1.0}%
\definecolor{LightGrey}{rgb}{0.975,0.975,0.975}
\lstdefinelanguage{balzac}{
	commentstyle=\color{Gray},
	morecomment=[l]{//},
	morecomment=[s]{/*}{*/},
	classoffset=0,
        escapechar=\$,
	morekeywords={const,transaction,input,output,absLock,relLock,key,network,package},
	keywordstyle=\color{TealBlue}\bfseries,
	classoffset=1,
	morekeywords={sig,versig,verscript,verrec,fun,unit,int,string,bool,address,uint,date,checkDate,sha256},
	keywordstyle=\color{Blue}\bfseries,
	classoffset=2,
	morekeywords={BTC,true,and,or,if,then,else,def},
	keywordstyle=\color{Plum}\bfseries,
	classoffset=3,
	morekeywords={arg,val,wit,out,in},
	keywordstyle=\color{MidnightBlue}\bfseries,
}

%% file: macro.tex
\newcommand{\ifempty}[3]{%
  \ifthenelse{\isempty{#1}}{#2}{#3}%
}

\newcommand{\ifdots}[3]{%
  \ifthenelse{\equal{#1}{...}}{#2}{#3}%
}

\newcommand{\hidden}[1]{}

\renewcommand{\vec}[1]{\boldsymbol{#1}}

\newcommand{\Real}[1]{\mathrm{Real}}

\newcommand{\codefont}{\fontsize{10}{10}\selectfont}

\newcommand{\lineno}[1]{{\tt\codefont {\textcolor{NavyBlue}{#1}}}}

\newcommand{\eg}{e.g.\@\xspace}
\newcommand{\ie}{i.e.\@\xspace}

\newcommand{\emptyseq}{\varepsilon}

\newcommand{\fst}{\mathit{fst}}
\newcommand{\snd}{\mathit{snd}}

  {}

\newcommand{\sig}[3][]{\mathit{sig}^{#1}_{#2}\ifempty{#3}{}{({#3})}}

\newcommand{\txvername}{{\txColor{\sf ver}}}

\newcommand{\txver}[6]{\ifempty{#3}{\ifempty{#4}{\txvername_{#1}({#2},{#5},{#6})}
		{\txvername_{#1}^{{#3},{#4}}({#2},{#5},{#6})}}
	{\txvername_{#1}^{{#3},{#4}}({#2},{#5},{#6})}}

\newcommand{\stdtxT}[2]{\txFmt{U}_{{#1}}^{\txColor{#2}}}

\newcommand{\BTC}{\textup{%
  \leavevmode
  \vtop{\offinterlineskip %\bfseries
    \setbox0=\hbox{B}%
    \setbox2=\hbox to\wd0{\hfil\hskip-.03em
    \vrule height .3ex width .15ex\hskip .08em
    \vrule height .3ex width .15ex\hfil}
    \vbox{\copy2\box0}\box2}}\xspace}

\def\pmvColor{\color{ForestGreen}}
\newcommand{\pmvFmt}[1]{{\pmvColor{\sf #1}}}

\newcommand{\pmv}[2][]{\pmvFmt{#2}_{\pmvColor{#1}}\xspace}
\newcommand{\pmvA}[1][]{\pmv[{#1}]{A}}
\newcommand{\pmvB}[1][]{\pmv[{#1}]{B}}

\def\txColor{\color{MidnightBlue}}
\newcommand{\txFmt}[1]{{\txColor{\sf #1}}}

\newcommand{\tx}[2][]{\txFmt{#2}_{\txColor{#1}}}
\newcommand{\txT}[1][]{\tx[#1]{T}}

\newcommand{\txTi}[1][]{\txFmt{T'_{\txColor{{#1}}}}}

\newcommand{\txf}[1][]{\tx[#1]{f}} % transaction field

\newcommand{\txIn}[2][]{{\textsf{in}}\ifempty{#1}{\ifempty{#2}{}{: {#2}}}{({#1})\ifempty{#2}{}{: {#2}}}}
\newcommand{\txWit}[2][]{{\textsf{wit}}\ifempty{#1}{\ifempty{#2}{}{: {#2}}}{({#1})\ifempty{#2}{}{: {#2}}}}
\newcommand{\txOut}[2][]{{\textsf{out}}\ifempty{#1}{\ifempty{#2}{}{: {#2}}}{({#1})\ifempty{#2}{}{: {#2}}}}
\newcommand{\txAfterAbs}[2][]{{\textsf{absLock}}\ifempty{#1}{\ifempty{#2}{}{: {#2}}}{({#1})\ifempty{#2}{}{: {#2}}}}
\newcommand{\txAfterRel}[2][]{{\textsf{relLock}}\ifempty{#1}{\ifempty{#2}{}{: {#2}}}{({#1})\ifempty{#2}{}{: {#2}}}}

\DeclareMathAlphabet{\mathbfsf}{\encodingdefault}{\sfdefault}{bx}{n}

\newcommand{\bcB}[1][]{{\mathbfsf{\txColor{B}}}_{#1}}

\newcommand{\bnfdef}{::=}
\newcommand{\bnfmid}{\;|\;}

\newcommand{\sem}[2][]{\mbox{\ensuremath{\llbracket{#2}\rrbracket_{#1}}}}

\newcommand{\setenum}[1]{\{#1\}}

\newcommand{\seqat}[2]{{#1}.{#2}}

\crefname{appendix}{appendix}{appendices}
\Crefname{appendix}{Appendix}{Appendices}
\crefname{notation}{notation}{notations}
\Crefname{notation}{Notation}{Notations}

\definecolor{LightGrey}{rgb}{0.95,0.95,0.95}
\definecolor{keyword}{HTML}{7F0055}

\newlength\replength
\newcommand\repfrac{.1}

\newcommand\rulewidth{.6pt}
\newcommand\tdashfill[1][\repfrac]{\cleaders\hbox to \replength{%
  \smash{\rule[\arraystretch\ht\strutbox]{\repfrac\replength}{\rulewidth}}}\hfill}

\newcommand\tdotfill[1][\repfrac]{\cleaders\hbox to \replength{%
  \smash{\raisebox{\arraystretch\dimexpr\ht\strutbox-.1ex\relax}{.}}}\hfill}

\newcommand{\var}[2][]{#2_{#1}} % variables
\newcommand{\varX}[1][]{\var[#1]{x}}

\newcommand{\const}[2][]{#2_{#1}} % constants

\newcommand{\constPK}[1][]{\const[{#1}]{pk}}
\newcommand{\constSK}[1][]{\const[{#1}]{sk}}

\newcommand{\constT}[1][]{\const[{#1}]{t}}

\newcommand{\val}[2][]{#2_{#1}} % values
\newcommand{\valV}[1][]{\val[#1]{v}}
\newcommand{\valVi}[1][]{\val[#1]{v'}}

\newcommand{\ArgA}[1][]{\vec{a}_{#1}}

\newcommand{\expe}{\ensuremath{e}}
\newcommand{\expei}{\ensuremath{e}'}

\newcommand{\versig}[2]{{\sf versig}({#1},{#2})}
\newcommand{\hashE}[1]{{\sf H}(#1)}
\newcommand{\hashSem}[1]{\ifempty{#1}{H}{H(#1)}}
\newcommand{\afterAbs}[2]{{\sf absAfter}~{#1}:{#2}}
\newcommand{\afterRel}[2]{{\sf relAfter}~{#1}:{#2}}
\newcommand{\op}[3]{\ensuremath{#2~\textsf{#1}~#3}}

\newcommand{\sizeE}[1]{\ensuremath{| #1 |}}
\newcommand{\ifE}[3]{\mathsf{if}~{#1}~\mathsf{then}~{#2}~\mathsf{else}~{#3}}
\newcommand{\ifSem}[3]{\mathit{if}~{#1}~\mathit{then}~{#2}~\mathit{else}~{#3}}
\newcommand{\andE}{~{\sf and}~}
\newcommand{\orE}{~{\sf or}~}

\newcommand{\rtx}{{\sf rtx}}
\newcommand{\rtxWit}{\rtx.\txWit[]{}}
\newcommand{\ctxo}[2]{{\sf ctxo}({#2})\ifempty{#1}{}{.{#1}}}
\newcommand{\rtxo}[2]{{\sf rtxo}({#2})\ifempty{#1}{}{.{#1}}}

\newcommand{\verscript}[2]{{\sf verscr}\ifempty{#1}{}{(#1, #2)}}
\newcommand{\verrec}[1]{{\sf verrec}\ifempty{#1}{}{(#1)}}

\newcommand{\txarg}{{\color{MidnightBlue}\sf arg}}
\newcommand{\txscript}{{\color{MidnightBlue}\sf scr}}
\newcommand{\txval}{{\color{MidnightBlue}\sf val}}
\newcommand{\outidx}{{\sf outidx}}
\newcommand{\inidx}{{\sf inidx}}

\newcommand{\sizeF}[1]{\mathit{size}\ifempty{#1}{}{({#1})}}
\newcommand{\true}{\mathit{true}}
\newcommand{\false}{\mathit{false}}

\def\contrColor{\color{RubineRed}}
\newcommand{\contrFmt}[1]{{\contrColor{\it #1}}}
\newcommand{\contrC}[1][]{\mathord{\contrFmt{C}_{\contrColor{#1}}}}
\newcommand{\contrCi}[1][]{\mathord{\contrC[#1]\contrColor{'}}}

\newcommand{\contrAdvC}[2]{\mathcal{C}} % computational contract advertisement

\newcommand{\revealname}{\textup{\texttt{reveal}}}
\newcommand{\wherename}{\textup{\texttt{if}}}

\newcommand{\splitname}{\textup{\texttt{split}}}
\newcommand{\splitC}[1]{\splitname \ifempty{#1}{}{\; #1}}
\newcommand{\splitB}[2]{{#1} \rightarrow {#2}}

\newcommand{\withdrawname}{\textup{\texttt{withdraw}}}
\newcommand{\withdrawC}[1]{\withdrawname\ifempty{#1}{}{\; {\pmv{#1}}}}

\newcommand{\authC}[2]{{\pmv{#1}}\,\textup{\texttt{:}}\,{#2}}

\newcommand{\afterName}{\texttt{after}}
\newcommand{\afterRelName}{\texttt{afterRel}}
\newcommand{\afterC}[2]{\textup{\afterName}\,{#1}\,\textup{\texttt{:}}\,{#2}}
\newcommand{\afterRelC}[2]{\textup{\afterRelName}\,{#1}\,\textup{\texttt{:}}\,{#2}}

\newcommand{\ifC}[1]{\,\wherename\, {#1} : }

\newcommand{\cVar}[2][]{\texttt{\textup{#2}}_{#1}}
\newcommand{\cVarX}[1][]{\cVar[#1]{X}}
\newcommand{\cVarY}[1][]{\cVar[#1]{Y}}
\newcommand{\cVarW}[1][]{\cVar[#1]{W}}

\newcommand{\call}[2]{{#1}({#2})}

\newcommand{\confContr}[3][]{\langle {#2}, {#3} \rangle_{#1}}
\newcommand{\confDep}[3][]{\langle {#2}, {#3} \rangle_{#1}}

\newenvironment{nscenter}
 {\parskip=0pt\par\nopagebreak\centering}
 {\parskip=2pt\par\noindent} % \ignorespacesafterend

%% file: abstract.tex
\begin{abstract}
  Covenants are linguistic primitives that extend the Bitcoin script language,
  allowing transactions to constrain the scripts of the redeeming ones.
  Advocated as a way of improving the expressiveness of Bitcoin contracts
  while preserving the simplicity of the UTXO design,
  various forms of covenants have been proposed over the years.
  A common drawback of the existing descriptions is the lack of formalization,
  making it difficult to reason about properties and supported use cases.
  In this paper we propose a formal model of covenants,
  which can be implemented with minor modifications to Bitcoin.
  We use our model to specify some complex Bitcoin contracts,
  and we discuss how to exploit covenants to design 
  high-level language primitives for Bitcoin contracts.
\end{abstract}

%% file: intro.tex
\section{Introduction}
\label{sec:intro}

Bitcoin is a decentralised infrastructure to transfer cryptocurrency between users.
The log of all the currency transactions is recorded in 
a public, append-only, distributed data structure, called blockchain.
Bitcoin implements a model of computation called \emph{Unspent Transaction Output (UTXO)}:
each transaction holds an amount of currency, and specifies conditions 
under which this amount can be redeemed by a subsequent transaction, which spends the old one.
Compared to the \emph{account-based} model, implemented \eg by Ethereum,
the UTXO model does not require a shared mutable state:
the current state is given just by the set of unspent transaction outputs on the blockchain.
While, on the one hand, this design choice fits well with the inherent concurrency of transactions,
on the other hand the lack of a shared mutable state substantially complicates
leveraging Bitcoin to implement \emph{contracts},
\ie protocols which transfer cryptocurrency according to programmable rules.

The literature has shown that Bitcoin contracts support a surprising variety of use cases, including \eg
crowdfunding~\cite{wikiassurance,bitcoinsok},
lotteries and other gambling games \cite{Andrychowicz14bw,BZ17bw,bitcoinsok,Bentov14crypto,Kumaresan15ccs,Miller16zerocollateral},
contingent payments~\cite{Banasik16esorics},
micro-payment channels~\cite{Poon15lighting,bitcoinsok},
and other kinds of fair computations~\cite{Andrychowicz14sp,Kumaresan14ccs}.
Despite this apparent richness, the fact is that Bitcoin contracts
cannot express most of the use cases that are mainstream in other blockchain platforms
(\eg, decentralised finance).
There are several factors that limit the expressiveness of Bitcoin contracts.
Among them, the crucial one is the script language used to express the redeeming conditions within transactions.
This language only features a limited set of logic, arithmetic, and cryptographic operators,
but its has no loops, and it cannot access parts of the spent and of the redeeming transaction.

Several extensions of the Bitcoin script language have been proposed,
with the aim to improve the expressiveness of Bitcoin contracts, while adhering to the UTXO model.
Among these extensions, \emph{covenants} are a class of script operators that allow a transaction 
to constrain how its funds can be used by the redeeming transactions.
Covenants may also be recursive, by requiring the script of the redeeming transaction to contain the same covenant
of the spent one.
As noted by~\cite{Oconnor17bw}, recursive covenants would allow to implement 
Bitcoin contracts that execute state machines, by appending transactions to trigger state transitions.

Although the first proposals of covenants date back at least to 2013~\cite{bitcointalk-covenants}, 
and that they are supported by Bitcoin fork ``Bitcoin Cash''~\cite{cashscript-covenants},
their inclusion into Bitcoin is still uncertain,
mainly because of the extremely cautious approach to implement changes to Bitcoin~\cite{BIP0002}.
Still, the emerging of Bitcoin layer-2 protocols, like \eg the Lightning Network~\cite{Poon15lighting},
has revived the interest in covenants,
as witnessed by a recent Bitcoin Improvement Proposal (BIP 119~\cite{BIP119,Swambo20bitcoin}),
and by the incorporation of covenants in Liquid's extensions to Bitcoin Script~\cite{Nick20liquid}.

Since the goal of the existing proposals 
is to show how implementing covenants would impact on the performance of Bitcoin,
they describe covenants from a low-level, technical perspective.
We believe that a proper abstraction and formalization of covenants would also be useful,
as it would simplify reasoning on the behaviour of Bitcoin contracts and on their properties.

\paragraph{Contributions}

We summarise our main contributions as follows:
\begin{itemize}
  
\item we introduce a formal model of Bitcoin covenants, 
  inspired by the informal, low-level presentation in~\cite{Moser16bw}.
  
\item we use our formal model to specify complex Bitcoin contracts, 
  which largely extend the set of use cases expressible in pure Bitcoin;
  
\item we discuss how to exploit covenants in the design of high-level language primitives
  for Bitcoin contracts.
  
\end{itemize}

%% file: bitcoin.tex
\section{The pure Bitcoin}
\label{sec:bitcoin}

We start by illustrating the Bitcoin transaction model.
To this purpose we adapt the formalization in~\cite{bitcointxm},
omitting the parts that are irrelevant for our subsequent technical development.

\paragraph{Transactions}

In its simplest form, a Bitcoin transaction
allows a user to transfer cryptocurrency (the \emph{bitcoins}, $\BTC$) to someone else.
For this to be possible, bitcoins must be created at first.
This is obtained through \emph{coinbase} transactions
(\ie, the first transaction of each mined block), whose typical form is:
\begin{nscenter}
  \small
  \begin{tabular}[t]{|l|}
    \hline
    \multicolumn{1}{|c|}{$\txT[0]$} \\
    \hline
    \txIn{$\bot$} \\
    \txWit{$\bot$} \\
    \txOut{$\{ \txscript: \versig{\constPK[\pmvA]}{\rtxWit}, \txval: 1 \BTC \}$} \\
    \hline
  \end{tabular}
\end{nscenter}

We identify $\txT[0]$ as a coinbase transaction by its $\txIn{}$ field, 
which does not point to any other previous transaction on the blockchain
(formally, we model this as the undefined value $\bot$).
The $\txOut{}$ field contains a pair,
whose first element is a \emph{script},
and the second one is the amount of bitcoins that will be redeemed 
by a subsequent transaction which points to $\txT[0]$ and satisfies its script.
In particular, the script $\versig{\constPK[\pmvA]}{\rtxWit}$ 
verifies the signature in the $\txWit{}$ field of the redeeming transaction ($\rtx$)
against $\pmvA$'s public key $\constPK[\pmvA]$.

Assume that $\txT[0]$ is on the blockchain, and that $\pmvA$ wants to transfer $1 \BTC$ to $\pmvB$.
To do this, $\pmvA$ can append to the blockchain a new transaction, \eg:
\begin{nscenter}
  \small
  \begin{tabular}[t]{|l|}
    \hline
    \multicolumn{1}{|c|}{$\txT[1]$} \\
    \hline
    \txIn{$\txT[0]$} \\
    \txWit{$\sig{\constSK[\pmvA]}{\txT[1]}$} \\
    \txOut{$\{ \txscript: \versig{\constPK[\pmvB]}{\rtxWit}, \txval: 1 \BTC \}$} \\
    \hline
  \end{tabular}
\end{nscenter}

The $\txIn{}$ field points to $\txT[0]$,
and the $\txWit{}$ field contains 
$\pmvA$'s signature on $\txT[1]$ (but for the $\txWit{}$ field itself).
This witness makes the script within $\txT[0].\txOut{}$ evaluate to true,
hence the redemption succeeds, and $\txT[0]$ is \emph{spent}.

The transactions $\txT[0]$ and $\txT[1]$ above only use part of the features of Bitcoin.
More in general, transactions can collect bitcoins from many inputs,
and split them between many outputs;
further, they can use more complex scripts, and specify time constraints.
Following the formalization in~\cite{bitcointxm}, 
we represent transactions as records with the following fields:
\begin{itemize}
\item $\txIn{}$ is the list of \emph{inputs}. 
  Each of these inputs is a \emph{transaction output} $(\txT,i)$, 
  referring to the $i$-th output field of $\txT$.
\item $\txWit{}$ is the list of \emph{witnesses},
  of the same length as the list of inputs.
  Intuitively, for each $(\txT,i)$ in the $\txIn{}$ field, 
  the witness at the same index must make the $i$-th output script 
  of $\txT$ evaluate to true.
\item $\txOut{}$ is the list of \emph{outputs}.  
  Each output is a record $\{ \txscript: \expe, \txval: \valV \}$, 
  where $\expe$ is a script, and $\valV$ is a currency value.
\item $\txAfterAbs{}$ is a value, indicating the first moment in time
  when the transaction can be added to the blockchain;
\item $\txAfterRel{}$ is a list of values, of the same length as the list of inputs.
  Intuitively, if the value at index $i$ is $n$, 
  the transaction can be appended to the blockchain only if
  at least $n$ time units have passed since the input transaction at index $i$
  has been appended.
\end{itemize} 

We let $\txf$ range over transaction fields, 
and we use the standard dot notation to access the fields of a record.
For a transaction output $(\txT,i)$ and $\txf \in \setenum{\txscript,\txval}$, 
we write $(\txT,i).\txf$ for $\txT.\txOut[i]{}.\txf$.
For uniformity, we assume that $\txAfterAbs{}$ is a list of unit length;
we omit null values in $\txAfterAbs{}$ and $\txAfterRel{}$.
When graphically rendering transactions, we usually write
$\txf(1) : \ell_1 \cdots \txf(n) : \ell_n$ for $\txf : \ell_1 \cdots \ell_n$, 
or just $\txf : \ell_1$ when $n=1$ (as in $\txT[0]$ and $\txT[1]$ above).
When clear from the context, we just write the name $\pmvA$ of a user in place
of her public/private keys, \eg we write $\versig{\pmvA}{\expe}$ for $\versig{\constPK[\pmvA]}{\expe}$,
and $\sig{\pmvA}{\txT}$ for $\sig{\constSK[\pmvA]}{\txT}$.

\paragraph{Scripts}
\label{sec:bitcoin-scripts}

Bitcoin scripts are small programs written in a non-Turing equivalent language.
Whoever provides a witness that makes the script evaluate to ``true'', 
can redeem the bitcoins retained in the associated (unspent) output.
In our model, scripts are terms with the following syntax,
where $\circ \in \setenum{+,-,=,<}$,
and where we write sequences of scripts in bold notation:
\[
\begin{array}{rcclclclclclc}
 \expe 
  & \; \bnfdef \;
  & \valV  
  & \bnfmid 
  & \expe \circ \expe 
  & \bnfmid 
  & \seqat{\vec{\expe}}{\expe}
  & \bnfmid 
  & \ifE{\expe}{\expe}{\expe}  
  & \bnfmid
  & \rtxWit
  & \bnfmid
  & 
  \\[2pt]
  & 
  & \sizeE{\expe} 
  & \bnfmid
  & \hashE{\expe} 
  & \bnfmid
  & \versig{\vec{\expe}}{\vec{\expe}}
  & \bnfmid
  & \afterAbs{\expe}{\expe} 
  & \bnfmid 
  & \afterRel{\expe}{\expe}
\end{array}
\]

Besides values $\valV$ and the basic arithmetic/logical operators, 
scripts feature operators to access the elements of a sequence
($\seqat{\vec{\expe}}{\expe}$),
to access the witnesses of the redeeming transaction ($\rtxWit$), 
to compute the size $\sizeE{\expe}$ of a bitstring
and its hash $\hashE{\expe}$.
The script $\versig{\vec{\expe}}{\vec{\expei}}$
evaluates to true iff the sequence of signatures 
resulting from the evaluation of $\vec{\expei}$ (say, of length $m$)
is verified by using $m$ out of the $n$ keys resulting from the evaluation of~$\vec{\expe}$.
The expressions \mbox{$\afterAbs{\expe}{\expei}$} and
\mbox{$\afterRel{\expe}{\expei}$} define absolute and relative time constraints:
they evaluate as $\expei$ if the constraints are satisfied,
otherwise their semantics is undefined.
We assume a basic type system which rules out ill-formed scripts.

We define in~\Cref{fig:bitcoin:scripts} the semantics of scripts.
The script evaluation function $\sem[\txT,i]{\cdot}$ takes two parameters:
$\txT$ is the redeeming transaction, and
$i$ is the index of the redeeming input/witness.
We denote with $\hashSem{}$ a public hash function, with $\sizeF{n}$ the size (in bytes) of an integer $n$,
and with $\txvername$ a multi-signature verification function 
(see~\cite{bitcointxm} for the definition of these semantic operators).
All the operators are \emph{strict},
\ie they evaluate to~$\bot$ if some of their operands is~$\bot$. 
We use syntactic sugar for scripts, 
\eg 
$\false$ denotes $\op{=}{1}{0}$,
$\true$ denotes $\op{=}{1}{1}$, while
$\expe \andE \expei$ denotes $\ifE{\expe}{\expei}{\false}$,
and $\expe \orE \expei$ denotes $\ifE{\expe}{\true}{\expei}$.

\begin{figure}[t]
  \[
  \begin{array}{c}
    \sem[\txT,i]{\valV} 
    = \valV 
    \qquad
    \sem[\txT,i]{\expe \circ \expe'} 
    = \sem[\txT,i]{\expe} \circ_\bot \sem[\txT,i]{\expe'}
      \quad 
      (\circ \in \{+, -, =, <\})
    \\[6pt]
    \sem[\txT,i]{\seqat{\vec{\expe}}{\expei}} = \sem[\txT,i]{\expe_j} \;\; 
    \text{if $\vec{\expe} = \expe_1 \cdots \expe_k$, $\sem[\txT,i]{\expei} = j$, and $1 \leq j \leq k$}
    \\[6pt]
    \sem[\txT,i]{\ifE{\expe_0}{\expe_1}{\expe_2}} 
    = \ifSem{\sem[\txT,i]{\expe_0}}{\sem[\txT,i]{\expe_1}}{\sem[\txT,i]{\expe_2}}
    \qquad
    \sem[\txT,i]{\rtxWit} = \txT.\txWit[i]{}
    \\[6pt]
    \sem[\txT,i]{\sizeE{\expe}} 
    = \sizeF{\sem[\txT,i]{\expe}}
    \qquad
    \sem[\txT,i]{\hashE{\expe}} 
    = \hashSem{\sem[\txT,i]{\expe}}
    \\[6pt]
    \sem[\txT,i]{\versig{\expe_1 \cdots \expe_n}{\expei_1 \cdots \expei_m}}
    = \txver{\scriptsize \sem[\txT,i]{\expe_1} \cdots \sem[\txT,i]{\expe_n}}{\sem[\txT,i]{\expei_1} \cdots \sem[\txT,i]{\expei_m}}{}{}{\txT}{i}
    \\[6pt]
    \sem[\txT,i]{\afterAbs{\expe}{\expei}} 
    = \ifSem{\txT.\txAfterAbs{} \geq \sem[\txT,i]{\expe}}{\sem[\txT,i]{\expei}}{\bot}
    \\[6pt]
    \;\;\, \sem[\txT,i]{\afterRel{\expe}{\expei}} 
    \, = \, \ifSem{\txT.\txAfterRel[i]{} \geq \sem[\txT,i]{\expe}}{\sem[\txT,i]{\expei}}{\bot}
  \end{array}
  \]
  \vspace{-10pt}
  \caption{Semantics of Bitcoin scripts.}
  \label{fig:bitcoin:scripts}
\end{figure}

\paragraph{Blockchains}

We model a blockchain $\bcB$ as a sequence of transactions $\txT[0] \cdots \txT[n]$.
For simplicity, we abstract from the fact that Bitcoin groups transactions into time-stamped blocks,
and we identify the time-stamp of a transaction with its position in the blockchain.
We say that the $j$-th output of the transaction $\txT[i]$ in the blockchain is \emph{spent}
iff there exists some transaction $\txT[i']$ in the blockchain (with $i' > i$) 
and some $j'$ such that $\txT[i'].\txIn[j']{} = (\txT[i],j)$.

A transaction $\txT[n]$ is \emph{valid} with respect to the blockchain $\bcB = \txT[0] \cdots \txT[n-1]$
whenever the following conditions hold:
\begin{enumerate}
  \item for each input $i$ of $\txT[n]$, if $\txT[n].\txIn[i]{} = (\txTi,j)$ then:
  \begin{enumerate}
  \item \label{consistent-update:match} 
    $\txTi = \txT[h]$, for some $h < n$ (\ie, $\txTi$ is one of the transactions in $\bcB$);
  \item \label{consistent-update:unspent}
    the $j$-th output of $\txT[h]$ is not spent in~$\bcB$;
  \item \label{consistent-update:output} 
    $\sem[{\txT[n]},i]{\txT[h].\txOut[j]{}} = \true$;
    \item \label{consistent-update:relLock} 
      $n-h \geq \txT[n].\txAfterRel[i]{}$; 
  \end{enumerate}
\item \label{consistent-update:absLock} 
  $n \geq \txT[n].\txAfterAbs{}$;
\item \label{consistent-update:value}
  the sum of the amounts of the inputs of $\txT[n]$ is greater or equal
  to the sum of the amount of its outputs
  (the difference between the amount of inputs and that of outputs is the \emph{fee} paid to miners).
\end{enumerate}

\noindent
The Bitcoin consensus protocol ensures that 
each $\txT[i]$ in the blockchain (except the coinbase $\txT[0]$) is valid
with respect to the sequence of past transactions $\txT[0] \cdots \txT[i-1]$.

%% file: covenants.tex
\section{Extending Bitcoin with covenants}
\label{sec:covenants}

To extend Bitcoin with covenants, we amend the transaction model of the previous section as follows:
\begin{itemize}
  \item we add to each output a field $\txarg: \ArgA$, 
    where $\ArgA$ is a sequence of values. 
    Intuitively, the extra field can be used to encode a state within transactions.
  \item we add script operators to access the outputs of the current transaction and of the redeeming one  
    (by contrast, pure Bitcoin scripts can only access the whole redeeming transaction, but not its parts).
    \bartnote{check!!}
  \item we add script operators to check whether the output scripts in the redeeming transaction 
    match a given script, or a given output of the current transaction
    (by contrast, in pure Bitcoin the redeeming transaction is only used when verifying signatures).
\end{itemize}

% We start by introducing some notation to access the parts of transaction outputs.
% Namely, we extend the range of the meta-variable $\txf$ to include $\txarg$, $\txscript$, and $\txval$,
% with the following meaning.
% If $\txT.\txOut[i]{} = (\ArgA,\expe,\valV)$, then 
% $\txT.\txarg(i) = \ArgA$, $\txT . \txscript(i) = \expe$, and $\txT . \txval(i) = \valV$.

\begin{figure}[t]
  \begin{align*}
    \expe \; 
    \bnfdef \;\;
    \cdots 
    & \bnfmid \;
    \ctxo{\txf}{\expe}
    && \text{access part of the current transaction}
    \\
    & \bnfmid \;
    \rtxo{\txf}{\expe}
    && \text{access part of the redeeming transaction}
    \\
    & \bnfmid \;
    \outidx
    && \text{index of the redeemed output}
    \\
    & \bnfmid \;
    \inidx
    && \text{index of the redeeming input}
    \\
    & \bnfmid \;
    \verscript{\expe}{\expe}
    && \text{covenant}
    \\
    & \bnfmid \;
    \verrec{\expe}
    && \text{recursive covenant}
  \end{align*}
  \vspace{-15pt}
  \caption{Extended Bitcoin scripts ($\txf \in \setenum{\txarg,\txscript,\txval}$).}
  \label{fig:covenants:scripts-syntax}
\end{figure}

We extend the syntax of scripts in~\Cref{fig:covenants:scripts-syntax},
and in~\Cref{fig:covenants:scripts-semantics} we define their semantics.
As in pure Bitcoin, the script evaluation function takes as parameters
the redeeming transaction $\txT$ and the index $i$ of the redeeming input/witness.
From them, it is possible to infer
the current transaction $\txT' = \fst(\txT.\txIn[i]{})$,
and the index $j = \snd(\txT.\txIn[i]{})$ of the redeemed output.
The script $\ctxo{\txf}{k}$ evaluates to the field $\txf$
of the $k$-th output of the current transaction;
similarly, $\rtxo{\txf}{k}$ operates on the redeeming transaction.
The symbols $\outidx$ and $\inidx$ evaluate, respectively,
to the index of the redeemed output and to that of the redeeming input.
The last two scripts specify covenants, in basic and recursive form.
The basic covenant $\verscript{k}{\expei}$ checks that the $k$-th output script of the redeeming transaction
is syntactically equal to $\expei$ (note that $\expei$ is \emph{not} evaluated).
The recursive covenant $\verrec{k}$ checks that the $k$-th output of the redeeming transaction
is syntactically equal to the redeemed output script.

\begin{figure}[t]
  \small
  \begin{align*}
    & 
      \sem[\txT,i]{\rtxo{\txf}{\expe}} 
      = (\txT,\sem[\txT,i]{\expe}) . \txf
    &&
      \sem[\txT,i]{\ctxo{\txf}{\expe}}
      = (\fst \, \txT.\txIn[i]{},\sem[\txT,i]{\expe}) . \txf
    \\[5pt]
    &
       \sem[\txT,i]{\inidx} 
       = i
    &&
      \sem[\txT,i]{\outidx} 
      = \snd \, \txT.\txIn[i]{}
    \\[5pt]
    & \sem[\txT,i]{\verscript{\expe}{\expei}} 
      = (\txT,\sem[\txT,i]{\expe}) . \txscript \equiv \expei
    && 
    \sem[\txT,i]{\verrec{\expe}} 
    = (\txT,\sem[\txT,i]{\expe}) . \txscript \equiv \txT.\txIn[i]{} . \txscript
  \end{align*}
  \vspace{-15pt}
  \caption{Semantics of extended scripts.}
  \label{fig:covenants:scripts-semantics}
\end{figure}

%% file: examples.tex
\section{Use cases}
\label{sec:examples}

We illustrate the expressive power of our extension through a series of use cases,
which, at the best of our knowledge, cannot be expressed in Bitcoin.
We denote with $\stdtxT{\pmvA}{\valV}$ an unspent transaction output 
$\{ \txarg: \emptyseq,\, \txscript: \versig{\pmvA}{\rtx.\txWit{}},\, \txval: \valV \BTC \}$,
where $\emptyseq$ denotes the empty sequence (we will usually omit $\txarg$ when empty).

%% file: crowdfunding.tex
\subsection{Crowdfunding}

\newcommand{\contrCF}{\contrFmt{CF}}

Assume that a start-up $\pmv{Z}$ wants to raise funds through a crowdfunding campaign. 
The target of the campaign is to gather at least $\valV \BTC$ by time $t$.
The contributors want the guarantee that if this target is not reached,
then they will get back their funds after the expiration date.
The start-up wants to ensure that 
contributions cannot be retracted before time $t$, or once $\valV \BTC$ have been gathered.

We implement this use case without covenants, but just constraining
the $\txval$ field of the redeeming transaction.
To fund the campaign, a contributor $\pmvA[i]$ publishes the transaction $\txT[i]$ 
in \Cref{fig:crowdfunding} (left),
which uses the following script:
\[
\contrCF 
\; = \;
\begin{array}{l}
  ( \versig{\pmv{Z}}{\rtxWit} \andE \rtxo{\txval}{1} \ge \valV ) \; \orE 
  \\[4pt]
  \afterAbs{t}{\versig{\pmvA[i]}{\rtxWit}}
\end{array}
\]
This script is a disjunction between two conditions.
The first condition allows $\pmv{Z}$ to redeem the bitcoins deposited in this output,
provided that the output at index 1 of the redeeming transaction pays at least $\valV\BTC$
(note that this constraint, rendered as $\rtxo{\txval}{1} \ge \valV$, 
is not expressible in pure Bitcoin).
The second condition allows $\pmvA[i]$ to get back her contribution after the expiration date $t$.

\begin{figure}[t]
  \begin{center}
    \resizebox{\textwidth}{!}{
      \begin{tabular}{|l|}
        \hline
        \multicolumn{1}{|c|}{$\txT[i]$} \\
        \hline
        \txIn{$\;\;\stdtxT{\pmvA[i]}{\valV[i]}$} \\
        \txWit{$\cdots$} \\
        \txOut{$\{ \txarg: \emptyseq,\, \txscript: \contrCF,\; \txval: \valV[i] \BTC \}$} \\
        \hline
      \end{tabular}
      \begin{tabular}{|l|}
        \hline
        \multicolumn{1}{|c|}{$\txT[\pmv{Z}]$} \\
        \hline
        \txIn{$(\txT[1], 1) \cdots (\txT[n], 1)$} \\
        \txWit{$\sig{\pmv{Z}}{\txT[\pmv{Z}]} \cdots \sig{\pmv{Z}}{\txT[\pmv{Z}]}$} \\
        \txOut{$\{ \txarg: \emptyseq,\, \txscript: \versig{\pmv{Z}}{\rtxWit},\, \txval: \valVi \BTC \}$} \\
        \hline
      \end{tabular}
    }
  \end{center}
  \vspace{-15pt}
  \caption{Transactions for the crowdfunding contract.}
  \label{fig:crowdfunding}
\end{figure}

Once contributors have deposited enough funds
(\ie, there are $n$ transactions $\txT[1], \ldots, \txT[n]$ 
with $\valVi = \valV[1] + \cdots \valV[n] \geq \valV$),
$\pmv{Z}$ can get $\valVi\BTC$ by appending $\txT[\pmv{Z}]$ to the blockchain.
Note that, compared to the assurance contract in the Bitcoin wiki~\cite{wikiassurance},
ours offers more protection to the start-up.
Indeed, while in~\cite{wikiassurance} any contributor can retract her funds at any time,
this is not possible here until time $t$.

%% file: non-fungible-token.tex
\subsection{Non-fungible tokens}
\label{sec:nft}

\newcommand{\nftoken}{\contrFmt{NFT}}
\newcommand{\nftokeni}{\contrFmt{NFT'}}

A non-fungible token represents the ownership of a physical or logical asset,
which can be transferred between users.
Unlike fungible tokens (\eg, ERC-20 tokens in Ethereum~\cite{erc20}), 
where each token unit is interchangeable with every other unit, 
non-fungible ones have unique identities.
Further, they do not support split and join operations, unlike fungible tokens.

We start by implementing a subtly flawed version of the non-fungible token.
Consider the transactions in~\Cref{fig:non-fungible-token}, which use the following script:
\[
\nftoken 
\; = \;
\versig{\ctxo{\txarg}{1}}{\rtxWit} \andE \verrec{1} \andE \rtxo{\txval}{1} = 1
\]

User $\pmvA$ mints a token by depositing $1 \BTC$ in $\txT[0]$:
to declare her ownership over the token, she sets $\txOut[1]{}.\txarg$ to her public key.
To transfer the token to $\pmvB$, $\pmvA$ appends the transaction $\txT[1]$, 
setting its $\txOut[1]{}.\txarg$ to $\pmvB$'s public key.

To spend $\txT[0]$, the transaction $\txT[1]$ must satisfy the conditions specified by the script $\nftoken$:
\begin{inlinelist}
\item the $\txWit{}$ field must contain the signature of the current owner; % $\sig{\pmvA}{}$
\item the script at index 1 must be equal to that at the same index in $\txT[0]$;
\item the output at index 1 must have $1 \BTC$ value, to preserve the integrity of the token.
\end{inlinelist}
Once $\txT[1]$ is on the blockchain, $\pmvB$ can transfer the token to another user, 
by appending a transaction which redeems $\txT[1]$.

\begin{figure}[t]
  \begin{center}
    % \resizebox{\textwidth}{!}{
      \begin{tabular}{|l|}
        \hline
        \multicolumn{1}{|c|}{$\txT[0]$} \\
        \hline
        \\[-10pt]
        \txIn{$\;\stdtxT{\pmvA}{1}$} \\
        \txWit{$\sig{\pmvA}{\txT[0]}$} \\
        \txOut{$\{ \txarg: \pmvA,\; \txscript: \nftoken,\; \txval: 1 \BTC \}$} \\
        \hline
      \end{tabular}
      \;
      \begin{tabular}{|l|}
	\hline
	\multicolumn{1}{|c|}{$\txT[1]$} \\
	\hline
        \\[-10pt]
	\txIn{$(\txT[0], 1)$} \\
	\txWit{$\sig{\pmvA}{\txT[1]}$} \\
	\txOut{$\{ \txarg: \pmvB,\; \txscript: \nftoken,\; \txval: 1 \BTC \}$} \\
	\hline
      \end{tabular}
      %
    % }
  \end{center}
  \vspace{-15pt}
  \caption{$\pmvA$ creates a token with $\txT[0]$, and transfers it to $\pmvB$ with $\txT[1]$.}
  \label{fig:non-fungible-token}
\end{figure}

\begin{figure}[t]
  \begin{center}
    \begin{tabular}{|l|}
      \hline
      \multicolumn{1}{|c|}{$\txT[2]$} \\
      \hline
      \txIn{$(\txT[\pmvA], 1) \; (\txTi[\pmvA], 1)$} \\
      \txWit{$\sig{\pmvA}{\txT[2]} \; \sig{\pmvA}{\txT[2]}$} \\
      \txOut[1]{$\{ \txarg:\pmvA,\; \txscript:\nftoken,\; \txval:1 \BTC \}$}
      \\
      \txOut[2]{$\{ \txarg: \emptyseq\, , \hspace{3pt} \txscript: \versig{\pmvA}{\rtxWit},\; \txval: 1 \BTC \}$}
      \\
      \hline
    \end{tabular}
  \end{center}
  \vspace{-15pt}
  \caption{$\pmvA$ exploits the flaw to destroy a token, redeeming its value.}
  \label{fig:non-fungible-token:attack}
\end{figure}

The script $\nftoken$ has a design flaw, already spotted in~\cite{Moser16bw}:
we show how $\pmvA$ can exploit this flaw in \Cref{fig:non-fungible-token:attack}.
Suppose we have two unspent transactions: $\txT[\pmvA]$ and $\txTi[\pmvA]$,
both representing a token owned by $\pmvA$ (in their first and only output). 
The transaction $\txT[2]$ can spend both of them, since it complies with all the validity conditions:
indeed, $\nftoken$ only constrains the script in the first output of the redeeming
transaction, while the other outputs are only subject to the standard validity conditions
(in particular, that the sum of their values does not exceed the value in input).
Actually, $\txT[2]$ destroys one of the two tokens, and removes the covenant from the other one.

To solve this issue, we can amend the $\nftoken$ script as follows:
\[
\nftokeni \; = \; \versig{\ctxo{\txarg}{\outidx}}{\rtxWit} \andE \verrec{\inidx} \andE \rtxo{\txval}{\inidx} = 1
\]
The amended script correctly handles the case of a transaction which uses different outputs 
to store different tokens.
$\nftokeni$ uses $\ctxo{\txarg}{\outidx}$, instead of $\ctxo{\txarg}{1}$ in $\nftoken$, 
to ensure that, when redeeming a given output, 
the signature of the owner of the token at \emph{that} output is checked.
Further, $\nftokeni$ uses $\verrec{\inidx}$, instead of $\verrec{1}$ in $\nftoken$, 
to ensure that the covenant is propagated exactly to the transaction output
which is redeeming that token (\ie, the one at index $\inidx$).
Notice that the amendment would make $\txT[2]$ invalid: 
indeed, the script in $\txTi[\pmvA].\txOut[1]{}$ would evaluate to false:
\begin{align*}
  \sem[{\txT[2],2}]{\nftokeni}
  & = \sem[{\txT[2],2}]{\verrec{\inidx}} \; \land \; \cdots 
  \\
  & = (\txT[2],\sem[{\txT[2],2}]{\inidx}).\txscript \equiv \txT[2].\txIn[2]{}.\txscript  \; \land \; \cdots 
  \\
  & = (\txT[2],2).\txscript \equiv (\txTi[\pmvA],1).\txscript \; \land \; \cdots 
  \\
  & = \versig{\pmvA}{\rtxWit} \equiv \nftokeni  \; \land \; \cdots 
  \\
  & = \false
\end{align*}

% After the amendment, a transaction redeeming two tokens must produce two tokens.

An alternative patch, 
originally proposed in \cite{Moser16bw}, is to add a unique identifier $id$ to each token,
\eg by amending the $\nftoken$ script as follows:
\[
\nftoken \andE id=id
\]
This allows to mint distinguishable tokens. 
For instance, if the tokens in $\txT[\pmvA]$ and $\txTi[\pmvA]$ are distinguishable,
$\txT[2]$ cannot redeem both of them.

%% file: vault.tex
\subsection{Vaults}
\label{sec:vaults}

\newcommand{\nfv}{\contrFmt{V}}
\newcommand{\nfs}{\contrFmt{S}}
\newcommand{\nfr}{\contrFmt{R}}

Transaction outputs are usually secured by cryptographic keys
(\eg through the script $\versig{\constPK[\pmvA]}{\rtx.\txWit[]{}}$).
Whoever knows the corresponding private key (\eg,  $\constSK[\pmvA]$) can redeem such an output:
in case of key theft, the legitimate owner is left without defence.
Vault transactions, introduced in~\cite{Moser16bw}, are a technique to mitigate this issue,
by allowing the legitimate owner to abort the transfer.

To create a vault, $\pmvA$ deposits $1 \BTC$ in a transaction $\txT[V]$ with the script $\nfv$:
\begin{align*}
  \nfv
  & =  \versig{\pmvA}{\rtxWit} \andE \verscript{1}{\nfs}
  \\
  \nfs 
  & = \big( \afterRel{\constT}{\versig{\ctxo{\txarg}{\outidx}}{\rtxWit}} \big)
  \orE \versig{\pmv{Ar}}{\rtxWit}
\end{align*}

The transaction $\txT[V]$ can be redeemed with the signature of $\pmvA$, but only 
by a \emph{de-vaulting} transaction like $\txT[S]$ in~\Cref{fig:vault}, which uses the script $\nfs$.
The output of the de-vaulting transaction $\txT[S]$ can be spent by the user set in its $\txarg$ field,
but only after a certain time $\constT$ (\eg, by the transaction $\txT$ in~\Cref{fig:vault}). 
Before time $\constT$, $\pmvA$ can cancel the transfer
by spending $\txT[S]$ with her recovery key $\pmv{Ar}$.

\begin{figure}[t]
  \begin{center}
    \resizebox{\textwidth}{!}{
    \begin{tabular}{|l|}
      \hline
      \multicolumn{1}{|c|}{$\txT[V]$} \\
      \hline
      \\[-7pt]
      \txIn{$\stdtxT{\pmvA}{1}$} \\[2pt]
      \txWit{$\cdots$} \\[2pt]
      \txOut{$\{ \txscript:\nfv, \txval:1 \BTC \}$} \\[2pt]
      \hline
    \end{tabular}
    \begin{tabular}{|l|}
      \hline
      \multicolumn{1}{|c|}{$\txT[S]$} \\
      \hline
      \\[-7pt]
      \txIn{$\txT[V]$} \\[2pt]
      \txWit{$\sig{\pmvA}{\txT[S]}$} \\[2pt]
      \txOut{$\{ \txarg: \pmvB, \txscript:\nfs, \txval:1 \BTC \}$} \\[2pt]
      \hline
    \end{tabular}
    \begin{tabular}{|l|}
      \hline
      \multicolumn{1}{|c|}{$\txT$} \\
      \hline
      \txIn{$\txT[S]$} \\
      \txWit{$\sig{\pmvB}{\txT}$} \\
      \txOut{$\{ \txscript:\versig{\pmvB}{\rtxWit}, \txval:1 \BTC)$} \\
      \txAfterRel{$\constT$} \\
      \hline
    \end{tabular}
    }
  \end{center}
  \vspace{-15pt}
  \caption{Transactions for the basic vault.}
  \label{fig:vault}
\end{figure}

\paragraph{A recursive vault}

The vault in~\Cref{fig:vault} has a potential issue, in that the recovery key may also be subject to theft.
Although this issue is mitigated by hardware wallets 
(and by the infrequent need to interact with the recovery key),
the vault modelled above does not discourage any attempt at stealing the key.

The issue can be solved by using a recursive covenant in the vault script $\nfr$:
\[
\begin{array}{ll}
  {\sf if} \;\seqat{\ctxo{\txarg}{1}}{1} = 0 
  & \textcolor{gray}{\text{// current state: vault}}
  \\ 
  \qquad {\sf then} \; \versig{\pmvA}{\rtxWit} \andE \verrec{1} \andE & \\
  \qquad \seqat{\rtxo{\txarg}{1}}{1}=1 
  & \textcolor{gray}{\text{// next state: de-vaulting}} \\
  {\sf else} \,\; (\afterRel{\constT}{\versig{\seqat{\ctxo{\txarg}{1}}{2}}{\rtxWit}}) \orE 
  & \textcolor{gray}{\text{// current state: de-vaulting}} \\
  \qquad \versig{\pmv{Ar}}{\rtxWit} \andE \verrec{1} \andE & \\
  \qquad \seqat{\rtxo{\txarg}{1}}{1}=0 
  & \textcolor{gray}{\text{// next state: vault}}
\end{array}
\]

In this version of the contract, 
the vault and de-vaulting transactions (in \Cref{fig:rec-vault}) have the same script.
The first element of the $\txarg$ sequence encodes the contract state
(0 models the vault state, and 1 the de-vaulting state),
while the second element is the user who can receive the bitcoin deposited in the vault.
The recovery key $\pmv{Ar}$ can only be used to append the re-vaulting transaction $\txT[R]$,
locking again the bitcoin into the vault.

Note that key theft becomes ineffective:
indeed, even if both keys are stolen, the thief cannot take control
of the bitcoin in the vault, as $\pmvA$ can keep re-vaulting. 

\begin{figure}[t]
  \begin{center}
    \resizebox{\textwidth}{!}{
    \begin{tabular}{|l|}
      \hline
      \multicolumn{1}{|c|}{$\txT[V]$} \\
      \hline
      \\[-9pt]
      \txIn{$\stdtxT{\pmvA}{1}$} \\
      \txWit{$\cdots$} \\
      \txOut{$\{ \txarg:0, \txscript:\nfr, \txval:1 \BTC \}$} \\
      \hline
    \end{tabular}
    \begin{tabular}{|l|}
      \hline
      \multicolumn{1}{|c|}{$\txT[S]$} \\
      \hline
      \\[-9pt]
      \txIn{$\txT[V]$} \\
      \txWit{$\sig{\pmvA}{\txT[S]}$} \\
      \txOut{$\{ \txarg:1 \pmvB, \txscript:\nfr, \txval:1 \BTC \}$} \\
      \hline
    \end{tabular}
    \begin{tabular}{|l|}
      \hline
      \multicolumn{1}{|c|}{$\txT[R]$} \\
      \hline
      \\[-9pt]
      \txIn{$\txT[S]$} \\
      \txWit{$\sig{\pmv{Ar}}{\txT[R]}$} \\
      \txOut{$\{ \txarg:0, \txscript:\nfr, \txval:1 \BTC \}$} \\
      \hline
    \end{tabular}
    }
  \end{center}
  \vspace{-15pt}
  \caption{Transactions for the recursive vault.}
  \label{fig:rec-vault}
\end{figure}

%% file: ponzi.tex
\subsection{A pyramid scheme}
\label{sec:ponzi}

\newcommand{\expPonzi}{\contrFmt{P}}
\newcommand{\expPonziX}{\contrFmt{X}}

Ponzi schemes are financial frauds which lure users under the promise of high profits,
but which actually repay them only with the investments of new users.
A pyramid scheme is a Ponzi scheme where the scheme creator recruits other investors, 
who in turn recruit other ones, and so on.
Unlike in Ethereum, where several Ponzi schemes have been implemented 
as smart contracts~\cite{Bartoletti20fgcs,Chen18www},
the limited expressive power of Bitcoin contract only allows for off-chain schemes~\cite{Vasek15fc}.

We design the first ``smart'' pyramid scheme in Bitcoin 
using the transactions in~\Cref{fig:ponzi}, where:
\begin{align*}
  \expPonzi
  & = \verscript{1}{\expPonziX} \andE \rtxo{\txarg}{1} = \ctxo{\txarg}{\outidx} \andE \rtxo{\txval}{1} = 2 
  \\ 
  & \hspace{10pt} \andE \verrec{2} \andE \verrec{3}
  \\
  \expPonziX
  & = \versig{\ctxo{\txarg}{\outidx}}{\rtxWit}
\end{align*}

To start the scheme, a user $\pmvA[0]$ deposits $1\BTC$ in the transaction $\txT[0]$
(we burn this bitcoin for uniformity, so that each user earns at most $1\BTC$ from the scheme).
To make a profit, $\pmvA[0]$ must convince other two users, say $\pmvA[1]$ and $\pmvA[2]$,
to join the scheme. 
This requires the cooperation of $\pmvA[1]$ and $\pmvA[2]$ to publish a transaction which redeems $\txT[0]$.
The script $\expPonzi$ ensures that this redeeming transaction has the form of $\txT[1]$ in~\Cref{fig:ponzi}, \ie
$\txOut[1]{}$ transfers $2\BTC$ to $\pmvA[0]$,
while the scripts in $\txOut[2]{}$ and $\txOut[3]{}$ ensure that the same behaviour is recursively 
applied to $\pmvA[1]$ and $\pmvA[2]$.

Overall, the contract ensures that, as long as new users join the scheme, each one earns $1\BTC$.
Of course, as in any Ponzi scheme, at a certain point it will no longer be possible to find new users,
so those at the leaves of the transaction tree will just lose their investment.

\begin{figure}[t]
  \begin{center}
      \begin{tabular}{|l|}
        \hline
        \multicolumn{1}{|c|}{$\txT[0]$} \\
        \hline
        \\[-8pt]
        \txIn{$\;\;\stdtxT{\pmvA[0]}{1}$} \\[2pt]
        \txWit{$\sig{\pmvA[0]}{\txT[0]}$} \\[2pt]
        \txOut{$\{ \txarg:\pmvA[0], \txscript:\expPonzi, \txval:0 \BTC \}$} \\[2pt]
        \hline
      \end{tabular}
      \;
      \begin{tabular}{|l|}
        \hline
        \multicolumn{1}{|c|}{$\txT[1]$} \\
        \hline
        \\[-8pt]
        \txIn{$\; \txT[0] \; \stdtxT{\pmvA[1]}{1} \; \stdtxT{\pmvA[2]}{1}$} \\[3pt]
        \txWit{$\bot \; \sig{\pmvA[1]}{\txT[1]} \; \sig{\pmvA[2]}{\txT[1]}$} \\[3pt]
        \txOut[1]{$\{ \txarg:\pmvA[0], \txscript:\expPonziX, \txval:2\BTC \}$} \\[3pt]
        \txOut[2]{$\{ \txarg:\pmvA[1], \txscript:\expPonzi, \txval:0\BTC \}$} \\[3pt]
        \txOut[3]{$\{ \txarg:\pmvA[2], \txscript:\expPonzi, \txval:0\BTC \}$} \\[2pt]
        \hline
      \end{tabular}
  \end{center}
  \vspace{-15pt}
  \caption{Transactions for the pyramid scheme.}
  \label{fig:ponzi}
\end{figure}

%% file: kotet.tex
\subsection{King of the Ether Throne}
\label{sec:kotet}

\newcommand{\expKotET}{\contrFmt{K}}
\newcommand{\expKotETX}{\contrFmt{X}}

\emph{King of the Ether Throne}~\cite{kotet} is an Ethereum contract,
which has been popular for a while around 2016, 
until a bug caused its funds to be frozen.
The contract is initiated by a user, who pays an entry fee $\valV[0]$ to become the ``king''.
Another user can usurp the throne by paying $\valV[1] = 1.5 \valV[0]$ fee to the old king,
and so on until new usurpers are available.
Of course this leads to an exponential growth of the fee needed to become king, 
so subsequent versions of the contract introduced mechanisms to make the current king
die if not ousted within a certain time.
Although the logic to distribute money substantially differs from that in~\Cref{sec:ponzi}, 
this is still an instance of Ponzi scheme,
since investors are only paid with the funds paid by later investors.

We implement the original version of the contract, 
fixing the multiplier to $2$ instead of $1.5$, since Bitcoin scripts do not support multiplication.
The contract uses the transactions in~\Cref{fig:kotet} for the first two kings, $\pmvA[0]$ and $\pmvA[1]$, 
where:
\begin{align*}
  \expKotET
  & =
    \verrec{1} \,\andE\,
    \rtxo{\txarg}{2} = \ctxo{\txarg}{1} \,\andE\, 
  \\
  & \hspace{14pt} \rtxo{\txval}{2} \geq \ctxo{\txval}{2} + \ctxo{\txval}{2} \,\andE \verscript{2}{\expKotETX} % send jackpot to old king
  \\
  \expKotETX
  & = \versig{\ctxo{\txarg}{2}}{\rtxWit}
\end{align*}

We use the $\txarg$ field in $\txOut[1]{}$ to record the new king, 
and that in $\txOut[2]{}$ for the old one.
The clause $\rtxo{\txarg}{2} = \ctxo{\txarg}{1}$ in $\expKotET$ 
preserves the old king in the redeeming transaction.
The clause $\rtxo{\txval}{2} \geq \ctxo{\txval}{2} + \ctxo{\txval}{2}$ 
ensures that his compensation is twice the value he paid.
Finally, $\verscript{}{}$ guarantees that the old king can redeem his compensation via $\txOut[2]{}$.

\begin{figure}[t]
  \begin{center}
    \resizebox{\textwidth}{!}{
      \begin{tabular}{|l|}
        \hline
        \multicolumn{1}{|c|}{$\txT[0]$} \\
        \hline
        \\[-8pt]
        \txIn{$\;\;\stdtxT{\pmvA[0]}{\valV[0]}$} \\[2pt]
        \txWit{$\sig{\pmvA[0]}{\txT[0]}$} \\[2pt]
        \txOut[1]{$\{ \txarg:\pmvA[0], \txscript:\expKotET, \txval:0 \BTC \}$} \\[2pt]
        \txOut[2]{$\{ \txarg:\pmvA[0], \txscript:\versig{\pmvA[0]}{\rtxWit}, \txval:\valV[0] \BTC \}$} \\[2pt]
        \hline
      \end{tabular}
      \begin{tabular}{|l|}
        \hline
        \multicolumn{1}{|c|}{$\txT[1]$} \\
        \hline
        \\[-8pt]
        \txIn{$\;\; (\txT[0], 1) \; \stdtxT{\pmvA[1]}{\valV[1]}$} \\[2pt]
        \txWit{$\bot \; \sig{\pmvA[1]}{\txT[1]}$} \\[2pt]
        \txOut[1]{$\{ \txarg:\pmvA[1], \txscript:\expKotET, \txval:0\BTC \}$} \\[2pt]
        \txOut[2]{$\{ \txarg:\pmvA[0], \txscript:\expKotETX, \txval:\valV[1]\BTC \}$} \\[2pt]
        \hline
      \end{tabular}
    }
  \end{center}
  \vspace{-15pt}
  \caption{Transactions for King of the Ether Throne.}
  \label{fig:kotet}
\end{figure}

%% file: implementation.tex
\section{Implementing covenants on Bitcoin}

We now discuss how to implement covenants in Bitcoin, and their computational overhead.
First, during the script verification, we need to access both the redeeming transaction
and the one containing the output being spent.
This can be implemented by adding a new data structure to store unspent or partially unspent 
transaction outputs,
and modifying the entries of the UTXO set to link each unspent output to the enclosing transaction.

The language primitives that check the redeeming transaction script, 
$\verscript{}{}$ and $\verrec{}$, can be implemented through an opcode similar to
${\sf CheckOutputVerify}$ described in~\cite{Moser16bw}.
While \cite{Moser16bw} uses placeholders to represent variable parts of the script,
\eg, $\versig{\texttt{<\textit{pubKey}>}}{\rtx.\txWit{}}$, 
we use operators to access the needed parts of a transaction, \eg, $\versig{\ctxo{\txarg}{1}}{\rtx.\txWit{}}$. 
Thus, to check if two scripts are the same we just need to compare their hashes,
while~\cite{Moser16bw} needs to instantiate the placeholders.
Similarly, we can use the hash of the script within $\verscript{}{}$.
The work \cite{Oconnor17bw} implements covenants without introducing operators 
to explicitly access the redeeming transaction.
Instead, they exploit the current implementation of {\sf versig},
which checks a signature on data that is build by implicitly accessing 
the redeeming transaction, to define a new operator {\sf CheckSigFromStack}.

The $\txarg$ part of each output can be stored at the beginning of the output script,
without altering the structure of pure Bitcoin transactions.
Similarly to the implementation of parameters in Balzac~\cite{bitcointxm,balzac},
the arguments are pushed on the alternative stack at the beginning of the script, 
then duplicated and copied in the main stack before the actual output script starts.
Note that arguments need to be discharged when hashing the script for $\verrec{}$/$\verscript{}{}$.
For this, it is enough to skip a known-length prefix of the script.

Even though the use cases in~\Cref{sec:examples} extensively use non-standard scripts,
they can be encoded as standard transactions using {\sf P2SH}~\cite{wikip2sh}, 
as done in \cite{bitcointxm,balzac}.
Crucially, the hash also covers the $\txarg$ field, which is therefore not malleable.

%% file: bitml-covenants.tex
\section{Using covenants in high-level contract languages}
\label{sec:bitml-covenants}

\newcommand{\CFGindent}{\hspace{8pt}}

\newcommand{\valC}{\textsf{val}}
\newcommand{\putCov}[2][]{\,?{#2}\ifempty{#1}{}{\,\,\wherename\, {#1}}}

\newcommand{\ifS}{{\sf if} \,}
\newcommand{\thenS}{\,{\sf then} \,}
\newcommand{\elseS}{{\sf else} \,}

As witnessed by the use cases in~\Cref{sec:examples}, 
crafting a contract at the level of Bitcoin transactions can be complex and error-prone.
To simplify this task, the work \cite{BZ18bitml} has introduced a high-level contract language,
called BitML, with a secure compiler to pure Bitcoin transactions.
BitML has primitives to $\withdrawC{}$ funds from a contract,
to $\splitC{}$ a contract (and its funds) into subcontracts,
to request the authorization from a participant $\pmvA$ before proceeding with a subcontract $\contrC$
(written $\authC{\pmvA}{\contrC}$),
to postpone the execution of $\contrC$ after a given time $\constT$
(written $\afterC{\constT}{\contrC}$),
to $\revealname$ committed secrets, and to branch between two contracts (written $\contrC + \contrCi$).
A recent paper~\cite{BMZ20coordination} extends BitML with a new primitive 
that allows participants to (consensually) renegotiate a contract, 
still keeping the ability to compile into pure Bitcoin.

Despite the variety of use cases shown in~\cite{bitmlracket,BCZ18isola},
BitML has known expressiveness limits, given by the requirement to have pure Bitcoin as its compilation target.
For instance, BitML cannot specify recursive contracts
(just as pure Bitcoin cannot),
unless all participants agree to perform the recursive call~\cite{BMZ20coordination}.
In this~\namecref{sec:bitml-covenants} we discuss how to improve the expressiveness of BitML,
assuming to use Bitcoin with covenants as compilation target.
We illustrate our point by a couple of examples,
postponing the formal treatment of this extended BitML and of its secure compilation to future work.

Covenants allow us to extend BitML with the construct:
\[
\putCov[b]{\vec{\varX}} .\; \call{\cVarX}{\vec{\varX}}
\]
Intuitively, the prefix $\putCov[b]{\vec{\varX}}$
can be fired whenever a participant provides a sequence of arguments $\vec{\varX}$
and makes the predicate $b$ true.
Once the prefix is fired, the contract proceeds as the continuation 
$\call{\cVarX}{\vec{\varX}}$,
which will reduce according to the equation defining $\cVarX$.

Using this construct, we can model the ``King of the Ether Throne'' contract of~\Cref{sec:kotet}
(started by $\pmvA$ with an investment of $1\BTC$) as $\call{\cVarX}{\pmvA,1}$, where:
\begin{align*}
  \call{\cVarX}{\pmv{a},v} 
  & \; = \; \putCov[\valC \geq 2 v]{\pmv{b}} .\; \call{\cVarY}{\pmv{a},\pmv{b},\valC}
  \\
  \call{\cVarY}{\pmv{a},\pmv{b},v} 
  & \; = \; \splitC{\big( \splitB{0}{\call{\cVarX}{\pmv{b},v}} \mid \splitB{v}{\withdrawC{a}} \big) }
\end{align*}
The contract $\call{\cVarX}{\pmv{a},\valV}$ models a state where $\pmv{a}$ is the current king,
and $\valV$ is his investment.
The guard $\valC \geq 2 v$ becomes true when some participant injects funds into the contract,
making its value ($\valC$) greater than $2\valV$. 
This participant can choose the value for $\pmv{b}$, \ie the new king.
The contract proceeds as $\call{\cVarY}{\pmv{a},\pmv{b},\valC}$,
which has two parallel branches. 
The first branch makes $\valC\,\BTC$ available to the old king;
the second branch has zero value, and it reboots the game, 
recording the new king $\pmv{b}$ and his investment.

A possible computation of $\pmvA$ starting the scheme with $1\BTC$ is the following,
where we represent a contract $\contrC$ storing $\valV\BTC$ as a term $\confContr{\contrC}{\valV\BTC}$:
\begin{align*}
  \confContr{\call{\cVarX}{\pmvA,-}}{1\BTC}
  & \xrightarrow{} \confContr{\call{\cVarY}{\pmvA,\pmvB,2}}{2\BTC}
  && \text{($\pmvB$ pays $2\BTC$ fee)}
  \\
  & \xrightarrow{} \confContr{\call{\cVarX}{\pmvB,2}}{0\BTC} \mid \confContr{\withdrawC{A}}{2\BTC}
  && \text{(contract splits)}
  \\
  & \xrightarrow{} \confContr{\call{\cVarX}{\pmvB,2}}{0\BTC} \mid \confDep{\pmvA}{2\BTC}
  && \text{($\pmvA$ redeems $2\BTC$)}
  \\
  & \xrightarrow{} \confContr{\call{\cVarY}{\pmvB,\pmv{C},4}}{4\BTC} \mid \confDep{\pmvA}{2\BTC}
  && \text{($\pmv{C}$ pays $4\BTC$ fee)}
  \\
  & \xrightarrow{} \confContr{\call{\cVarX}{\pmv{C},4}}{0\BTC} \mid \confContr{\withdrawC{B}}{4\BTC} \mid \confDep{\pmvA}{2\BTC}
  && \text{(contract splits)}
  \\
  & \xrightarrow{} \confContr{\call{\cVarX}{\pmv{C},4}}{0\BTC} \mid \confDep{\pmvB}{4\BTC} \mid \confDep{\pmvA}{2\BTC}
  && \text{($\pmvB$ redeems $4\BTC$)}
\end{align*}

\begin{figure}[t]
  \centering
  \scalebox{1}{
\begin{lstlisting}[language=balzac,numbers=left,numbersep=5pt,xleftmargin=10pt,classoffset=1,morekeywords={},classoffset=2,morekeywords={},classoffset=3,morekeywords={q,oldK,newK,v},framexbottommargin=0pt]
def arg.1 = q     // state    0 = <X(A,-),1>
                  // state    1 = <Y(a,b,v),v>
                  // state    2 = <X(b,v),0> | <withdraw a,v>    
def arg.2 = oldK  // old King
def arg.3 = newK  // new king
def arg.4 = v     // paid fee

verrec(1) and                                  // out(1) preserves covenant
if ctxo(1).q = 0 then                          // state 0
      rtxo(1).q = 1                            // state transition 0 -> 1
  and rtxo(1).oldK = ctxo(1).newK              // usurp the throne
  and rtxo(1).val >= ctxo(1).val + ctxo(1).val // fee at least doubled
  and rtxo(1).v = rtxo(1).val                  // instantiate v    
else if ctxo(1).q = 1 then                     // state 1
      rtxo(1).q = 2                            // state transition 1 -> 2
  and rtxo(1).newK = ctxo(1).newK              // preserve new king
  and rtxo(1).v = ctxo(1).v                    // preserve v
  and rtxo(1).val = 0                          // reset value in out(1)	
  and rtxo(2).oldK = ctxo(1).oldK              // set old king
  and verscr(2,versig(ctxo(2).oldK,rtx.wit))   // covenant to pay old king
  and rtxo(2).val = ctxo(1).val                // preserve value in out(2)
else if ctxo(1).q = 2 then                     // state 2
      rtxo(1).q = 1                            // state transition 2 -> 1
  and rtxo(1).oldK = ctxo(1).newK              // usurp the throne
  and rtxo(1).val >= ctxo(1).v + ctxo(1).v     // fee at least doubled
  and rtxo(1).v = rtxo(1).val                  // update v
\end{lstlisting}
  }
  \caption{Script for King of the Ether Throne, obtained by compiling BitML.}
  \label{fig:kotet:bitml}
\end{figure}

Executing a step of the BitML contract corresponds, in Bitcoin, to appending a transaction
containing in $\txOut[1]{}$ the script in~\Cref{fig:kotet:bitml}. 
The script implements a state machine, using $\seqat{\txarg}{1}$ to record the current state,
and the other parts of $\txarg$ for the old king, the new king, and $v$.
The $\verrec{1}$ at line~\lineno{8} preserves the script in $\txOut[1]{}$.
To pay the old king, we use the $\verscript{}{}$ at line~\lineno{20},
which constrains the script in $\txOut[2]{}$ of the transaction
corresponding to the BitML state 
$\confContr{\call{\cVarX}{\pmv{b},v}}{0\BTC} \mid \confContr{\withdrawC{a}}{\valV\BTC}$.

We now apply our extended BitML to specify a more challenging use case, \ie
a recursive coin-flipping game where two players $\pmvA$ and $\pmvB$ repeatedly flip coins,
and the one who wins two consecutive flips takes the pot.
The precondition to stipulate the contract requires each player to deposit $1\BTC$ as a bet.
The game first makes each player commit to a secret, using a timed-commitment protocol~\cite{Boneh00crypto}.
The secrets are then revealed, and the winner of a flip is determined as a function of the two secrets.
The game starts another flip if the current winner is different from that of the previous flip,
otherwise the pot is transferred to the winner.

We model the recursive coin-flipping game as the (extended) BitML contract $\call{\cVarX[\pmvA]}{\pmv{C}}$, 
where $\pmv{C} \neq \pmvA,\pmvB$, using the following defining equations:
\[
\begin{array}{rl}
  % 
  % A: ?: X_commB(W,#1) + afterRel t: wd B
  %
  \call{\cVarX[\pmvA]}{\pmv{w}} 
  & \; = \; \authC{\pmvA}{\putCov{h_{\pmvA}} \, . \, \call{\cVarX[\pmvB]}{\pmv{w},h_{\pmvA}}} 
  \; + \; \afterRelC{t}{\withdrawC{\pmvB}}
  \\[4pt]
  % 
  % B: ?: X_revA(W,a,#1) + afterRel t: wd A 
  % 
  \call{\cVarX[\pmvB]}{\pmv{w},h_{\pmvA}} 
  & \; = \; \authC{\pmvB}{\putCov{h_{\pmvB}} \, . \, \call{\cVarY[\pmvA]}{\pmv{w},h_{\pmvA},h_{\pmvB}}} 
  \; + \; \afterRelC{t}{\withdrawC{\pmvA}}
  \\[4pt]
  % 
  % X_revA(W,a,b) = ?pa if H(pa) = a . X_revB(W,pa,b) afterRel t: wd B
  % 
  \call{\cVarY[\pmvA]}{\pmv{w},h_{\pmvA},h_{\pmvB}} 
  & \; = \; \putCov[H(s_{\pmvA}) = h_{\pmvA}]{s_{\pmvA}} \, . \, \call{\cVarY[\pmvB]}{\pmv{w},s_{\pmvA},h_{\pmvB}}
  \\[4pt]
  & \CFGindent
  \; + \; \afterRelC{t}{\withdrawC{\pmvB}}
  \\[4pt]
  % 
  % X_revB(W,pa,b) = // TODO 0<=b<=1 da aggiungere
  % ?pb if H(pb)=b 
  % 
  \call{\cVarY[\pmvB]}{\pmv{w},s_{\pmvA},h_{\pmvB}} 
  & \; = \; \putCov[H(s_{\pmvB}) = h_{\pmvB} \andE 0 \leq s_{\pmvB} \leq 1]{s_{\pmvB}} \, . \, \call{\cVarW}{\pmv{w},s_{\pmvA},s_{\pmvB}}
  \\[4pt]
  & \CFGindent
  \; + \; \afterRelC{t}{\withdrawC{\pmvA}}
  \\[4pt]
  %
  % pa=pb and W=A . wd A
  % if a=b and W/=A . X_commA(A)
  % if a/=b and W=B . wd B
  % if a/=b and W/=B . X_commA(B)
  % afterRel t: wd A
  %
  \call{\cVarW}{\pmv{w},s_{\pmvA},s_{\pmvB}} 
  & \; = \; \ifC{s_{\pmvA} = s_{\pmvB} \andE \pmv{w}=\pmvA}{\withdrawC{\pmvA}}
    \hspace{20pt} \textcolor{gray}{\text{// $\pmvA$ won twice}}
  \\[4pt]
  & \CFGindent
  + \ifC{s_{\pmvA} = s_{\pmvB} \andE \pmv{w}\neq\pmvA} \call{\cVarX[\pmvA]}{\pmvA} 
    \hspace{46pt} \textcolor{gray}{\text{// $\pmvA$ won last flip}}
  \\[4pt]
  & \CFGindent
  + \ifC{s_{\pmvA} \neq s_{\pmvB} \andE \pmv{w}=\pmvB}{\withdrawC{\pmvB}}
    \hspace{20pt} \textcolor{gray}{\text{// $\pmvB$ won twice}}
  \\[4pt]
  & \CFGindent
  + \ifC{s_{\pmvA} \neq s_{\pmvB} \andE \pmv{w}\neq\pmvB} \call{\cVarX[\pmvA]}{\pmvB} 
    \hspace{46pt} \textcolor{gray}{\text{// $\pmvB$ won last flip}}
\end{array}
\]

The contract $\call{\cVarX[\pmvA]}{\pmv{w}}$ models a state where $\pmv{w}$ is the last winner,
and $\pmvA$ must commit to her secret.
To do that, $\pmvA$ must authorize an input $h_{\pmvA}$, which represents the hash of her secret.
If $\pmvA$ does not commit within $t$, then the pot can be redeemed by $\pmvB$ as a compensation
(here, the primitive $\afterRelC{t}{\contrC}$ models a relative timeout).
Similarly, $\call{\cVarX[\pmvB]}{\pmv{w}}$ models $\pmvB$'s turn to commit.
In $\call{\cVarY[\pmvA]}{\pmv{w},h_{\pmvA},h_{\pmvB}}$, $\pmvA$ must reveal her secret $s_{\pmvA}$, 
or otherwise lose her deposit.
The contract $\call{\cVarY[\pmvB]}{\pmv{w},s_{\pmvA},h_{\pmvB}}$ is the same for $\pmvB$,
except that here we additionally check that $\pmvB$'s secret is either $0$ or $1$
(this is needed to ensure fairness, as in the two-player lottery in~\cite{BZ18bitml}).
The flip winner is $\pmvA$
if the secrets of $\pmvA$ and $\pmvB$ are equal, 
otherwise it is $\pmvB$.
If the winner is the same as the previous round, the winner can withdraw the pot, 
otherwise the game restarts, recording the last winner.

This coin flipping game is fair, \ie the expected payoff of a \emph{rational} player
is always non-negative, notwithstanding the behaviour of the other player.

%% file: conclusions.tex
\section{Conclusions and future work}
\label{sec:conclusions}

We have proposed a formalisation of Bitcoin covenants,
and we have exploited it to present a series of use cases which appear to be unfeasible in pure Bitcoin.
We have introduced high-level contract primitives that exploit covenants
to enable recursion, and allow contracts to receive new funds and parameters at runtime.

\paragraph{Known limitations}
Most of the scripts crafted in our use cases would produce non-standard transactions,
that are rejected by Bitcoin nodes.
To produce standard transactions from non-standard scripts, we can exploit {\sf P2SH}~\cite{wikip2sh}.
This requires the transaction output to commit to the hash of the script,
while the actual script is revealed in the witness of the redeeming transaction.
Since, to check its hash, the script needs to be pushed to the stack, 
and  the maximum size of a stack element is 520 bytes, longer scripts would be rejected.
This clearly affects the expressiveness of contracts, as already observed in~\cite{bitmlracket}.
In particular, since the size of a script grows with the number 
of contract states (see \eg \Cref{fig:kotet:bitml}),
contracts with many states would easily violate the 520 bytes limit.
The introduction of Taproot \cite{bip341} would mitigate this limit.
For scripts with multiple disjoint branches, 
Taproot allows the witness of the redeeming transaction
to reveal just the needed branch.
Therefore, the 520 bytes limit would apply to branches, instead of the whole script.
Another expressiveness limit derives from the fact that 
covenants can only constrain the scripts of the redeeming transaction.
While this is enough to express non-fungible tokens (see~\Cref{sec:nft}),
fungible ones seem to require more powerful mechanisms, because of the join operation.
An alternative technique to enhancing covenants is to implement fungible tokens 
natively~\cite{Chakravarty20utxoma,Chakravarty20tokens},
or to enforce their logic through a sidechain~\cite{Nick20liquid}.

\paragraph{Verification}
Although designing contracts in the UTXO model seems to be less error-prone than in the shared memory model, 
\eg because of the absence of reentrancy vulnerabilities 
(like the one exploited in the Ethereum DAO attack~\cite{DAO}), 
Bitcoin contracts may still contain security flaws.
Therefore, it is important to devise verification techniques to detect security issues that may lead
to the theft or freezing of funds.
Recursive covenants make this task harder than in pure Bitcoin, 
since they can encode infinite-state transition systems, as in most of our use cases.
Hence, model-checking techniques based on the exploration of the whole state space, 
like the one used in~\cite{Andrychowicz14formats}, cannot be applied.

\paragraph{High-level Bitcoin contracts}
The compiler of our extension of BitML is just sketched in~\Cref{sec:bitml-covenants},
and we leave as future work its formal definition,
as well as the extension of the computational soundness results of~\cite{BZ18bitml},
ensuring the correspondence between the symbolic semantics of BitML and 
the underlying computational level of Bitcoin.
Continuing along this line of research, it would be interesting
to study new linguistic primitives that fully exploit the expressiveness of Bitcoin covenants,
and to extend accordingly the verification technique of~\cite{BZ19post}.
Note that our extension of the UTXO model is more restrictive than the one in~\cite{Chakravarty20wtsc},
as the latter abstracts from the script language, just assuming that scripts
denote any pure functions~\cite{Zahnentferner18utxo}.
This added flexibility can be exploited to design expressive high-level contract languages
like Marlowe~\cite{Thompson18isola} and Plutus~\cite{Brunjes20arxiv}.

%% file: ack.tex
\paragraph{Acknowledgements}
Massimo Bartoletti is partially supported by Aut.\ Reg.\ of Sardinia project ``Sardcoin''.
Stefano Lande is partially supported by P.O.R.\ F.S.E.\ 2014-2020.
Roberto Zunino is partially supported by MIUR PON 2018 
``Distributed Ledgers for Secure Open Communities'' ARS01\_00587.